%% file: main.tex
\documentclass[preprint,12pt,3p,onecolumn,nonatbib]{elsarticle}
\usepackage[utf8]{inputenc}
\usepackage[T1]{fontenc}

\makeatletter 
\let\c@author\relax

\makeatother

\usepackage[toc=true,nogroupskip]{glossaries-extra}
\setabbreviationstyle{long-short}
\loadglsentries{niels-glossary}

\graphicspath{{./img/}}

\usepackage[
		hyperref=true,
		style=numeric,
		natbib=true,
		url=false
	]{biblatex}
\usepackage[hidelinks, pdftex]{hyperref}
\usepackage{multirow}
\usepackage{amssymb}
\usepackage{cleveref}

\journal{Nuclear Physics A}

\begin{document}

\begin{frontmatter}

\title{Implementation of Cluster expansion for hot QCD matter}

\author[1]{Niels-Uwe Friedrich Bastian}
\author[2]{Pasi Huovinen}
\author[3]{Elizaveta Nazarova}

\address[1]{University of Wrocław, pl Maksa Borna 9, 50-204 Wrocław, Poland}
\address[2]{Incubator of Scientific Excellence---Centre for Simulations of Superdense Fluids, University of Wrocław, pl Maksa Borna 9, 50-204 Wrocław, Poland}
\address[3]{Joint Institute for Nuclear Research, 141980 Dubna, Russia}

\begin{abstract}
We present a cluster expansion EoS model for strongly-interacting matter based on the generalized Beth-Uhlenbeck formalism to describe hadrons as bound clusters of quarks. This formalism can describe both confined and deconfined phases.
Our emphasis is on the region of vanishing baryon densities, where numerical solutions available from Lattice QCD predict a smooth crossover transition from hadron to quark matter.
Medium effects are taken into account as self energies, which are motivated from both perturbative QCD calculations and phenomenological models.
Parameters are tuned to Lattice QCD data and result in a good agreement of the thermodynamics.
\end{abstract}

\end{frontmatter}

\section{Introduction}
\label{sec:introduction}
The \gls{eos} of strongly-interacting matter is a subject of active investigation within nuclear and high-energy physics communities. In the region of phase diagram, which corresponds to high temperatures ($T$) and vanishing baryo-chemical potential ($\mu$), the Lattice Quantum Chromodynamics (LQCD) can be used, predicting a crossover transition between hadronic matter and a phase of deconfined quarks (\gls{qgp})\cite{Bazavov:2018mes}. Still, due to the known sign problem \cite{deForcrand:2009zkb}, \gls{lqcd} is not applicable in the region of finite densities (potentials). On the other hand, \gls{pqcd} can describe matter at asymptotically high densities, predicting a phase of deconfined quark matter. However, the region of phase diagram ``in between'' remains for the most part unknown. This in turn means, that the question about the location of transition between hadron and quark matter, as well as the order of this transition, remains open. In this regard there are two possibilities: either somewhere at finite $\mu$ there exists at least one \gls{cep}, denoting the change from crossover to first-order phase transition, or, alternatively, the crossover phase transition is present through the entire phase diagram, and the \gls{cep} does not exist. 

It should be noted, that several non-pertubative approaches to the description of \gls{qcd} phase diagram in the regimes of finite $T$ and $\mu$ are being developed over the last decades. For example, the Dyson-Schwinger equation approach \cite{Roberts:2000aa} has achieved remarkable progress towards a unified \gls{eos} of quark-hadron matter \cite{Fischer:2018sdj}. The phase diagram with a \gls{cep} could be obtained using the  Polyakov–Quark–Meson model \cite{Schaefer:2007pw,Steinheimer:2013xxa}, and recently this approach has included the formulation of hadronization (describing hadrons as bound states of quarks) \cite{Alkofer:2018guy}. However, both of these approaches lack baryons in their description. Finally, another systematic non-pertubative approach based on the application of the functional renormalization group methods is being developed \cite{Pawlowski:2005xe}.

The concept of the model used in this work is outlined in \cite{Bastian:2018wfl}. 
In this approach the thermodynamics of a dense system of hadronic states (including their dissociation into the \gls{qgp}) is described on the basis of the underlying quark dynamics (taking into account that hadrons are bound states of quarks):
The generalized Beth-Uhlenbeck \gls{eos} that uses phase shifts in order to describe correlations and their modifications in a hot and dense environment is used for consistent description of bound and scattering states.
To overcome the limitations of the standard Beth-Uhlenbeck approach, fully dressed quasi-particle propagators are used. Moreover, the self energies for the quasi particle properties are being modelled by a \gls{rdf} approach, which is capable to treat even such intricate effects as confinement and chiral symmetry breaking ($\chi$SB) (as well as their medium dependence).

In \cite{Bastian:2018mmc} the model was applied to high net-baryon densities and low temperatures, showing its feasibility to describe a crossover or a first-order phase transition ending in a \gls{cep} at higher temperatures.
Therefore, the model may be applicable to the entire phase diagram of \gls{qcd}.
We test the validity of this approach by implementing it to high temperatures and low densities probed in ultrarelativistic \gls{hic}, and comparing to the results of \gls{lqcd} calculations.

The paper is structured as follows.
\Cref{sec:formalism} provides a detailed description of the general model of cluster expansion of strongly-interacting matter, followed by the specifics of the implementation to the regime of high temperatures and vanishing densities in \cref{sec:selfenergies}.
\Cref{sec:parametrisation} gives the details about the parametrisation realized for the high temperature region and the transition area.
Finally, \cref{sec:discussion} covers discussion of the results and further development of the model.

\section{Formalism}
\label{sec:formalism}

The cluster expansion model for strongly interacting matter is described in Refs.~\cite{Bastian:2018mmc,Bastian:2018wfl,Ropke:2012qv}.
The idea behind it is to describe matter, that consists of colored and colorless particles, utilizing the generalized Beth-Uhlenbeck approach for the description of a cluster expansion of strongly-correlated quark matter, where the clusters represent hadrons with spectral properties (bound states of quarks). The model then will describe thermodynamics of strongly-interacting matter, where partons are taken as quasi particles and hadrons as composites of these partons.

To avoid confusion we will follow a convention, in which the index $j$ can assume elementary partons (quarks and gluons), the index $i$ compound hadrons (mesons and baryons) and the index $l$ are all particles in the system (partons and hadrons).
In a cluster expansion, the particle density of elementary partons $j = \{u,d,s,g\}$ can be expressed as sum%
\begin{align}
	n_{j} = \sum_{l} A_{l,j} n_{l} (T,\mu)
\end{align}
over all species $l$ in the system (here: partons and hadrons), while for all fermionic contributions (quarks and baryons) their anti-particles are included implicitly.
The matrix $A_{l,j}$ is formed by the number of constituents $j$ in the cluster $l$.
In the generalized Beth-Uhlenbeck approach, the parton densities can be written as quasi particles
\begin{align}
	n_{l = \{u,d,s,g\}} (T, \mu) &= n_\mathrm{id} (T, g_l, M_l, \tilde\mu_l)\,,
\end{align}
with the ideal Fermi ($+$) or Bose ($-$) gas expressions for particle density
\begin{align}
	n_\mathrm{id} (T, g_l, M_l, \tilde\mu_l) &= g_{j} \int \frac{\mathrm{d}^{3}p}{(2 \pi)^{3}} \frac{1}{\mathrm e^{(\sqrt{p^{2} + M_j^{2}} + V_j - \mu_j)/T} \pm 1}\, ,
	\label{egn:density}
\end{align}
the degeneracy factor $g_l$, the effective mass $M_l$ and the effective chemical potential $\mu_l$.
The density of multi-particle states can be expressed by the generalized Beth-Uhlenbeck formula
\begin{align}
	n_{i} (T, \mu) = g_{i} \int \dfrac{\mathrm{d}^{3}p \mathrm{d}E}{(2 \pi)^{4}} f_{i} (E) 2 \sin^{2} \delta_{i} (E) \dfrac{\mathrm{d} \delta_{i} (E)}{\mathrm{d} E},
\end{align}
where the phase shift $\delta_{i} (E)$ is a medium-dependent quantity, which includes all properties of the particle species $i$, corresponding to its spectrum of bound and scattering states.
The species $i$ has degeneracy factor $g_{i}$ and obeys the Fermi ($+$) or Bose ($-$) distribution $f_{i} (E) = (\exp{[(E - \mu_{i})/T]} \pm 1)^{-1}$, respectively.
After substituting the integration over energy $\mathrm dE$ by an integration over effective mass $\mathrm dM$, using the quasi-particle dispersion relation $E_{i} = \sqrt{p^{2} + (M_{i})^{2}} + V_{i}$, we choose to follow a simple ansatz for the phase shifts  (they can only attain values of $n \pi$):
\begin{align}
	\delta_{i} (M) &= \pi \Theta (M - M_{i}) \Theta (M_{i}^{thr} - M)
\end{align}
where $M_{i} = m_{i} + S_{i}$ is the effective mass, $S_{i} (V_{i})$ are scalar (vector) self energies and $M_{i}^{thr}$ is the threshold mass, defined as the sum of constituent masses of a compound particle (hadron). 
Following this approach we get for the (bound) hadrons the expression
\begin{align}
\begin{split}
	n_i &= g_{i} \int \frac{\mathrm{d}^{3}p}{(2 \pi)^{3}} \left[ \frac{1}{\mathrm e^{(\sqrt{p^{2} + M_i^{2}} + V_i - \mu_i)/T} \pm 1} - \frac{1}{\mathrm e^{(\sqrt{p^{2} + (M^\mathrm{thr}_i)^{2}} + V_i - \mu_i)/T} \pm 1} \right] \Theta (M_{i}^\mathrm{thr} - M_{i}) \\
	&= (n_\mathrm{id} (T, g_i, M_i, \tilde\mu_i) - n_\mathrm{id} (T, g_i, M^\mathrm{thr}_i, \tilde\mu_i)) \Theta (M_{i}^\mathrm{thr} - M_{i})
\end{split}
\label{eqn:bounddensity}
\end{align}
which contains one usual ''free'' contribution, subtracted by another contribution, based on the mass of the multi-quark continuum threshold $M_{i}^{thr}$.
Analogously the expressions for scalar density and entropy density can be derived with their ideal gas expressions:
\begin{align}
    n^\mathrm S_\mathrm{id} (T, g_j, M_j, \tilde\mu_j) &= g_{j} \int \frac{\mathrm{d}^{3}p}{(2 \pi)^{3}} \frac{M_i}{\sqrt{p^{2} + M_j^{2}}} \frac{1}{\mathrm e^{(\sqrt{p^{2} + M_j^{2}} - \tilde\mu_j)/T} + 1}\,,
	\label{eqn:scalardensity}\\
    s_\mathrm{id} (T, g_j, M_j, \tilde\mu_j) &= g_{j} \int \frac{\mathrm{d}^{3}p}{(2 \pi)^{3}} \frac{\frac 43 p^2 + M_i^2}{T \sqrt{p^{2} + M_j^{2}}} \frac{1}{\mathrm e^{(\sqrt{p^{2} + M_j^{2}} - \tilde\mu_j)/T} + 1}\,.
	\label{eqn:entropy}
\end{align}

In order to implement our model to the region of high temperature and vanishing densities (and compare with \gls{lqcd}), we are including up, down, strange quarks and gluons as partons and considering all hadrons in the \gls{hrg} as clusters.
Now, the total scalar (S) and vector density (V) of the parton $j$ can be expressed as
\begin{align}
	n_j^\mathrm{(S,V)} (T, \{\mu_k\}) &= \sum_i A_{ij} n_i^\mathrm{(S,V)} (T, \mu_i) + n_{j,\mathrm{unbound}}^\mathrm{(S,V)} (T, \mu_j)\,.
\end{align}
Here $A_{ij}$ is the amount of partons $j\in \{u,d,s, g\}$ contained in the hadron $i \in \mathrm{HRG}$.
The effective quark mass $M_j = m_j + S_j$ is based on the bare mass $m_j$ and its scalar self energy $S_j$ and the earlier introduced threshold mass $M_i^\mathrm{thr} = \sum_j A_{ij} M_j$ of a hadron depends on the effective masses of its quark content.
In a similar way we can obtain the expression for the total entropy density of the system
\begin{align}
	s (T, \{\mu_k\}) &= \sum_i A_{ij} s_i (T, \mu_i) + s_{j,\mathrm{unbound}} (T, \mu_j)
\end{align}
and then the pressure as
\begin{align}
	p (T, \{\mu_k\}) &= \int_0^T \mathrm dT' s (T', \{\mu_k\}).
\end{align}
If one assumes that the self energies are derived as \gls{rdf} in analogy to \cite{Kaltenborn:2017hus}, the pressure would take the form
\begin{align}
	p (T, \{\mu_k\}) &= \sum_{i\in\mathrm{HRG}} p_i^\mathrm H (T, \mu_i) +  \sum_{j=\{u,d,s\}} p_j^\mathrm Q (T, \mu_j) + p_\mathrm g (T, \mu_\mathrm g) + \Theta\;.
\end{align}
Unfortunately, a consistent cluster mean field formulation for the \gls{rdf} approach is still an open problem, and therefore the corresponding expressions for the self energies $S_{j}$ and the term $\Theta$ are not known.

Therefore, in our work we utilise the approach of \cite{Typel:2018cap, Typel:2020ozc}.
Within this approach every self energy is expressed by the sum of a primary energy shift and a rearrangement contribution
\begin{align}
    S_j = \Delta m_j + m_j^\mathrm R\;.\label{eqn:selfenergydevided}
\end{align}
While the primary energy shift $\Delta m_j$ can be chosen almost arbitrarily, the rearrangement contributions $ m_j^\mathrm R$ ensure of thermodynamic consistency and depend on the primary shifts of all particle species:
\begin{align}
    m^\mathrm{R}_i &= \sum_j n^\mathrm{s}_j \frac{\partial \Delta m_j}{\partial n^\mathrm{s}_i}
\end{align}

Now we have all the basic tools at hand.
However, a primary shift for every particle type in the system needs to be assigned.

\section{Functional form of self energies}
\label{sec:selfenergies}
In the region of vanishing particle density it is sufficient to define scalar self energies since there are no contributions to the vector self energy. However, in future those will need to be introduced in order to derive and discuss higher baryon susceptibilities.

In order to create a model, where particles' self energies are consistently sensitive to each other, we introduce a generating density, which counts the presence of color charges based on the partial scalar densities (irrespective of its particular color): 
\begin{align}
    n_\mathrm s &= \sum_{i\in\mathrm{HRG}} A_i n_{\mathrm s,i} + \sum_{j=\{u,d,s\}} n_{\mathrm s,j} + 2  n_{\mathrm s,g}.
\end{align}
Here baryons contain three color charges ($A_i=3$), while mesons and gluons contain two color charges ($A_i=2$).

We concentrate first on the high temperature limit, where hadronic contributions can be neglected.
Here we can utilize results from pQCD in order to deduce the functional behaviour of the partonic masses at high temperatures.
As the next step, we discuss the inclusion of confinement with the help of linear string potential.
Last, but not least, we introduce necessary corrections for hadrons.

\subsection{Asymptotic limit}

At high temperatures we can assume, that all hadrons have disappeared and only quarks and gluons remain in the system, while asymptotically reaching the values or \gls{pqcd}.
In \cite{Laine:2006cp} an expression for the pressure, which can be considered applicable at sufficiently high temperatures, was derived. It can be shown (see \cref{sec:laine}), that the leading terms of this expression can be reformulated in quasi-particle terms as:
\begin{align}
\label{eqn:Laine}
    p/T^4 &\sim d_A \frac{\pi^2}{45} + \sum_{i=u,d,s} \left(p^\mathrm{qu}_i + A n^\mathrm{S}_i + B (n^\mathrm{S}_i)^2\right)\,.
\end{align}
Here the first term is the Stefan-Boltzmann limit of gluons and $A,B$ are non-trivial parameters, which are left open here, as we are only utilizing the functional form of the result.

The same behaviour can be reproduced in a \gls{rdf} approach (for example described in \cite{Bastian:2020unt}) using the generating functional
\begin{align}
    U &\sim  n_\mathrm{s} + n_\mathrm{s}^2\,,
\end{align}
leading to a following expression for scalar self energy of the particles:
\begin{align}
    S &\sim \mathrm{const} + n_\mathrm{s}\,.
\end{align}
Therefore, the highest order for scalar self energy is linear in scalar density.
It should be noted, that this diverging self energy does not contradict the Stefan-Boltzmann limit, because the thermal contributions are rising faster at $T\rightarrow\infty$.

\subsection{String potential and confinement}

At moderate temperatures non-perturbative contributions appear (e.g. confinement), which are up to now not fully understood and can not be analytically derived from \gls{qcd}.
In this work we are choosing a phenomenological approach, inspired by the linear string potential, which is discussed, for example, in \cite{Li:2015ida,Kaltenborn:2017hus,Bastian:2020unt}.
Accordingly, we introduce the following contributions to the scalar self energy of quarks and gluons:
\begin{align}
    S &\sim C n_\mathrm s^{1/3} + D n_\mathrm s^{-1/3}\;,
\end{align}
where $C$ is the coefficient for the one-gluon exchange and $D$ is the effective coefficient of linear string tension, which can be density-dependent in order to take into account color saturation \cite{Ropke:1986qs}.
The second term is diverging at low scalar densities (and hence low temperatures), leading to a statistical confinement of quarks and gluons due to their diverging masses.

\subsection{Hadronic corrections}

At high temperatures quarks and gluons should dominate due to their small masses.
In particular, the Heaviside function in \cref{eqn:bounddensity} will invoke a Mott transition once the threshold mass drops below the mass of the bound state.
On the other hand, the earlier introduced linear term for the quark masses will negate this effect, because quark masses will rise again over hadronic masses and an unphysical rehadronisation would occur.

In order to deal with this caveat, a linear contribution for hadrons $S \sim n_\mathrm{s}$ needs to be introduced, but to not change the behaviour at low temperatures or contradict the knowledge about hadronic in-medium effects in astrophysics, it should be folded with a form factor, resulting in:
\begin{align}
    S &\sim n_\mathrm s \frac{1}{\exp[(B_0 - n_\mathrm s)/B] - 1}\;.
\end{align}
This contribution would allow the hadrons to grow mass at sufficiently high temperatures, in order to suppress the unphysical rehadronisation, but does not affect lower temperatures.
It also does not interfere at all with applications at zero temperature and high density, because the scalar density in this region is comparably low.

\subsection{Resulting particle properties and self energies}
Now we are ready to write full expressions for particle properties, utilizing the assumptions, described in the previous subsections.
The resulting primary shifts for \cref{eqn:selfenergydevided} are:
\begin{align}
    \Delta m_{j = \{u,d,s, g\}} &= \sigma_{2,j}
        + \sigma_{4,j} n_\mathrm s
        + C_j n_\mathrm s^{1/3}
        + D_j n_\mathrm s^{-1/3} \mathrm e^{-\beta n_\mathrm s^2}
    \label{eqn:partonshift}
\end{align}
for partons, and
\begin{align}
    \Delta m_\mathrm{HRG} &= A n_\mathrm s \frac{1}{\exp[(B_0 - n_\mathrm s)/B] - 1}\;.
\end{align}
for hadrons.

The corresponding rearrangement contributions for each particle species are:
\begin{align}
    m^\mathrm R_{j = \{u,d,s\}}
        &= m^\mathrm R\;,\\
    m^\mathrm R_{g}
        &= 2 m^\mathrm R\;,\\
    m^\mathrm R_{i\in\mathrm{HRG}}
        &= A_i m^\mathrm R
\end{align}
with the common term
\begin{align}
\begin{split}
    m^\mathrm R =
        &\sum_{j = \{u,d,s, g\}} \left(\sigma_{4,j} + C_j n_\mathrm s^{-2/3} + D_j n_\mathrm s^{-4/3} \mathrm e^{-\beta n_\mathrm s^2} + D_j n_\mathrm s^{-1/3} \mathrm e^{-\beta n_\mathrm s^2}(-2\beta n_\mathrm s)\right) n_{\mathrm s,j}\\
        &+ \left(\frac{A}{\exp[(B_0 - n_\mathrm s)/B] - 1} +  \frac{A n_\mathrm s/B }{\exp[(n_\mathrm s - B_0)/B] - 1}\right) n_{\mathrm s, h}\;.
    \end{split}
\end{align}
Here $n_{\mathrm s,h}$ denotes the sum of scalar densities of all hadrons.

The arising coefficients $\sigma_{2,j}, \sigma_{4,j}, C_j, D_j, \beta, A, B$ and $B_0$ are open parameters, subject to a fit and will be discussed in the following section.
It will be shown, that the number of parameters can be drastically reduced, because many species dependencies are not necessary.

\section{Parametrisation}
\label{sec:parametrisation}

We aim to create a model, whose thermodynamics are comparable to the results of \gls{lqcd} in \cite{Bazavov:2014pvz}.
In order to optimize the fit procedure and high-temperature adjustments for the model, we need an analytic parametrisation, that provides us data in the whole temperature range.
First, we considered to use the fit of \cite{Parotto:2018pwx}, because it also provides parametrisations for higher susceptibilities, which will be important in our future work.
Unfortunately, it has unphysical behaviour at high temperatures, and therefore we are using in this work the fit from  \cite{Huovinen:2009yb}.

It proved to be effective to separate the region of high temperatures in which the system can be assumed to have only elementary partons and use it to fit only parameters which affect partons at high temperatures.
Afterwards those parameters are being fixed and the remaining ones are obtained by fitting the transition region, including all possible species.

\subsection{High temperature parametrisation}
Besides the absence of bound hadrons, the high temperature region has the advantage, that the confinement contributions do not play a role.
Instead of fitting our model here directly to the \gls{lqcd} thermodynamics, we use the effective masses of another quasi-particle approach for \gls{qgp} \cite{Mykhaylova:2019wci}.
This model describes the system of gluons ($g$), quarks ($u$, $d$ and $s$)  and their antiparticles with the effective quasi-particle masses dependent on the dynamically generated self-energies $\Pi_{i}$ as:
\begin{align}
    M_{i}^{2} = m_{i}^{2} + \Pi_{i} \, ,
\end{align}
where
\begin{align}
   &\Pi_i (T) = a_i \left(m_i \sqrt{\dfrac{G(T)^{2}}{6} T^{2}} + \dfrac{G(T)^{2}}{6} T^{2}\right),
\end{align}
taking the bare masses of particles as $m_{g} = 0~\mathrm{MeV}$, $m_{i=\{u,d\}} = 5~\mathrm{MeV}$, $m_{s} = 95~\mathrm{MeV}$, while the coefficients are $a_g = \left(3 + \frac{N_{f}}{2}\right) = \frac 92$ and $a_{i = \{u,d,s\}} = 2$.
Now the entropy density of the system can be computed by summing up the expression of eq~\eqref{eqn:entropy}
\begin{align}
    s = \sum_{j = u,d,s,g} s_\mathrm{id} (T, g_j, M_j, 0)\,.
    \label{eqn:entropysum}
\end{align}
Note, that the perturbative couplings have been replaced by an effective coupling $G(T)$, which in the high-temperature regime resembles the perturbative coupling for thermal momenta.
This model can be modified to describe pure Yang-Mills thermodynamics by setting the number of flavours equal to the spin-degeneracy factors of quarks and therefore introduces additional constraints for our model parameters.
In the setup above the only unknown quantity is $G(T)$, which can be obtained by inverting \cref{eqn:entropysum} with the help of aforementioned data for entropy density.
The resulting effective masses can now be fitted using our analytical approach for the scalar self energy.

We are concentrating on high temperatures (above $500~\mathrm{MeV}$) and neglect the confinement contribution in \cref{eqn:partonshift} as low-temperature term, to obtain the high-temperature behaviour of the primary mass shift:
\begin{align}
    \Delta m_{j = \{u,d,s,g\}} &= \sigma_{2,j}
        + \sigma_{4,j} n_\mathrm s
        + C_{j} n_\mathrm s^{1/3}
    \label{eqn:partonshifthight}
\end{align}
with the rearrangement term
\begin{align}
    m^\mathrm R = \sum_j \left(\sigma_{4,j} + C_j n_\mathrm s^{-2/3}\right) n_{\mathrm s,j},
\end{align}
which goes into \cref{eqn:selfenergydevided} as before.

\begin{figure}[t]
	\includegraphics[scale=0.7]{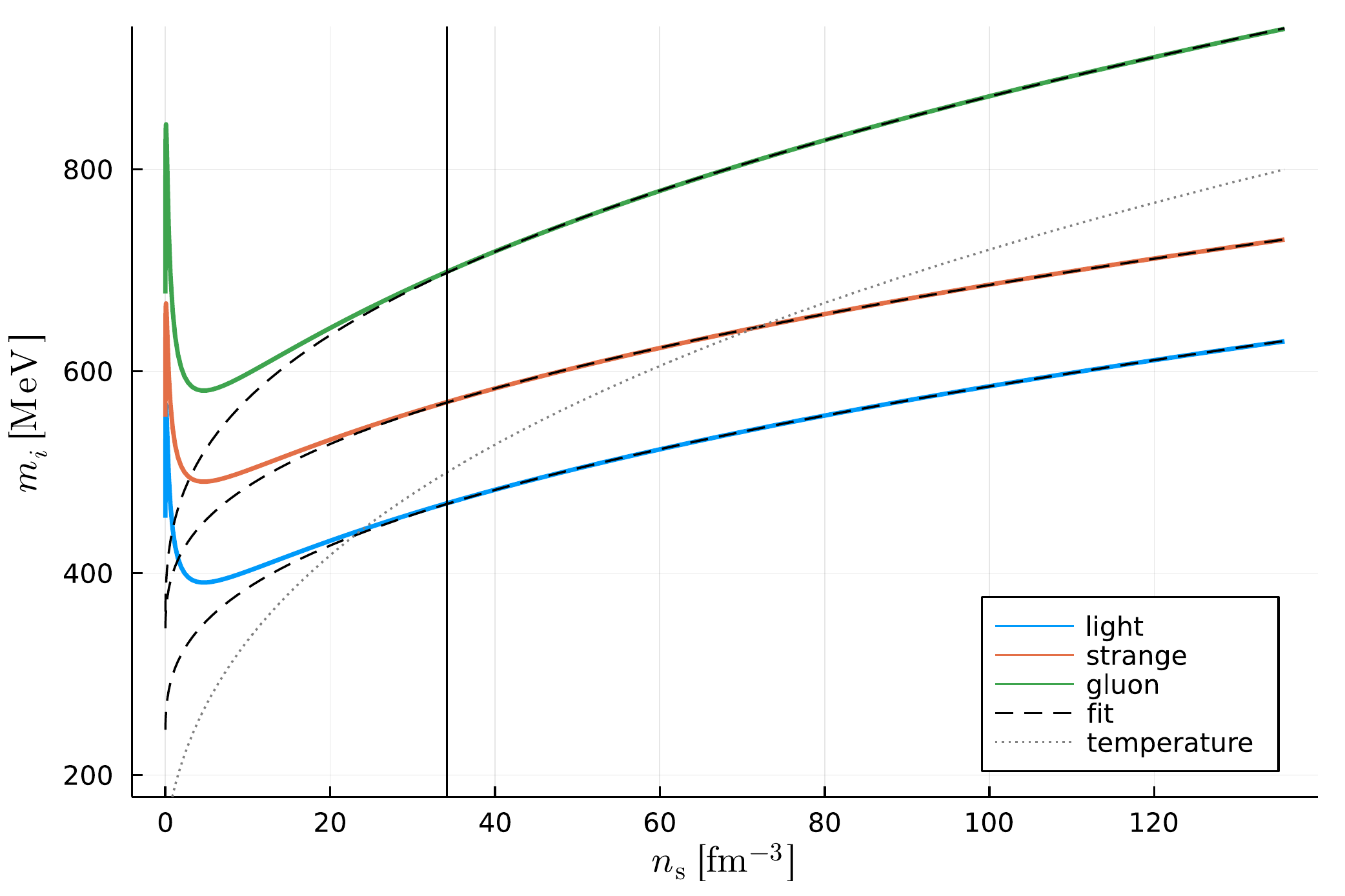}
	\caption{Mass fit for high temperatures ($T \ge 500MeV$), at which hadrons can be neglected. Vertical dotted line is threshold from which the curve is fitted. Dash-dotted line shows temperature.}
	\label{fig:highT_masses}
\end{figure}

\Cref{fig:highT_masses} shows the results of this fit as the dependency of masses on the scalar density, and the extracted parameters are:
\begin{align}\begin{split}
    &\sigma_{2,l} = 226.7~\mathrm{MeV},\quad
    \sigma_{2,s} = 182.2~\mathrm{MeV},\quad
    \sigma_{2,g} = 342.2~\mathrm{MeV},\\
    &\sigma_4 = 0.1414~\mathrm{MeV~fm^3},\quad
    C = 35.02~\mathrm{MeV~fm}.
\end{split}\end{align}
One can see that our functional in \cref{eqn:partonshifthight} reproduces excellently the masses of a model, which has been derived in a completely different way.
It is worth noting, that the parameters $\sigma_4$ and $C$ are species-independent.

\subsection{Parametrisation of the transition area}

Now that we have obtained the parameters $\sigma_{2,j}$, $\sigma_{4}$ and $C$ from the high-temperature fit, the remaining parameters are only the confinement parameters $D$, $D_g$ and $\beta$ and the hadronic corrections $A$, $B$ and $B_0$.
Them we obtain by fitting to the \gls{lqcd} thermodynamic data directly.
The fit has been performed using the entropy density, while ensuring to have reasonable particle fractions.
It turns out, that the best result could be achieved with vanishing color saturation $\beta = 0$.
The best fit values of the remaining parameters are

\begin{align}\begin{split}
    &A = 7.0~\mathrm{MeV~fm^3},\quad
    B = 2.8~\mathrm{fm^{-3}},\quad
    B_0 = 7.6~\mathrm{fm^{-3}},\\
    &D = 200.0~\mathrm{MeV~fm},\quad
    D_g = 100.0~\mathrm{MeV~fm}.
\end{split}\end{align}

\section{Discussion}
\label{sec:discussion}

\begin{figure}[t!]
	\includegraphics[scale=0.7]{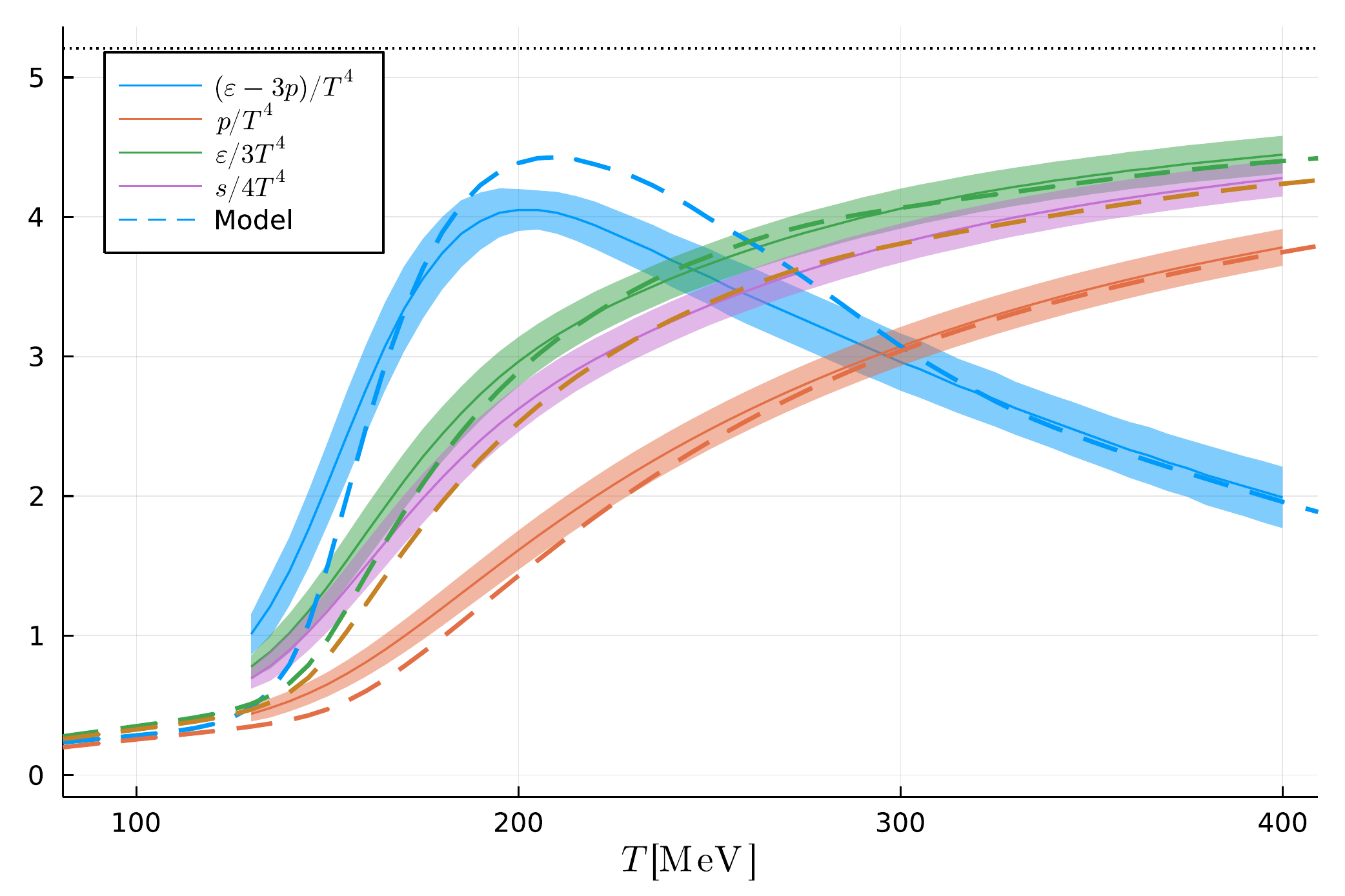}
	\caption{Thermodynamic observables of the model (dashed), compared to \gls{lqcd} data (shaded bands). The presented observables are the trace anomaly $\varepsilon - 3p$, pressure $p$, energy density $\varepsilon$ and entropy density $s$. The dotted line on top is the Stefan-Boltzmann limit of $p$, $\varepsilon$, and $s$.}
	\label{fig:thermodynamics}
\end{figure}

\begin{figure}[t]
	\includegraphics[scale=0.7]{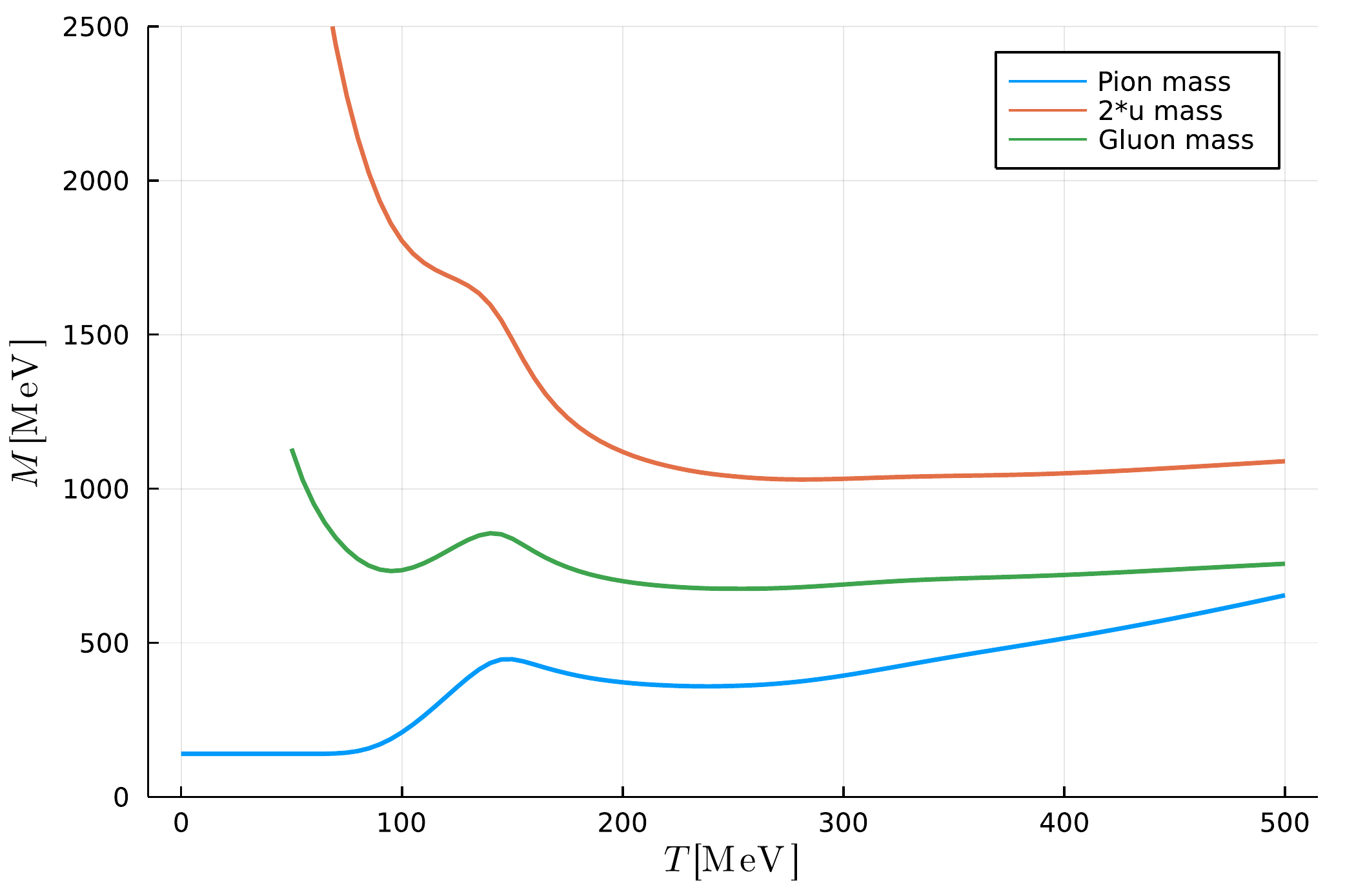}
	\caption{
	    Effective masses of pions, two light quarks and gluons. The mott transition, shown as crossing point between pions and two light quarks is outside the picture at $T \sim 600~\mathrm{MeV}$.
	    Both the quark and the gluon masses diverge at low temperatures, but the gluon line is cut due to numerical reasons.
	}
	\label{fig:masses}
\end{figure}

\begin{figure}
	\includegraphics[scale=0.7]{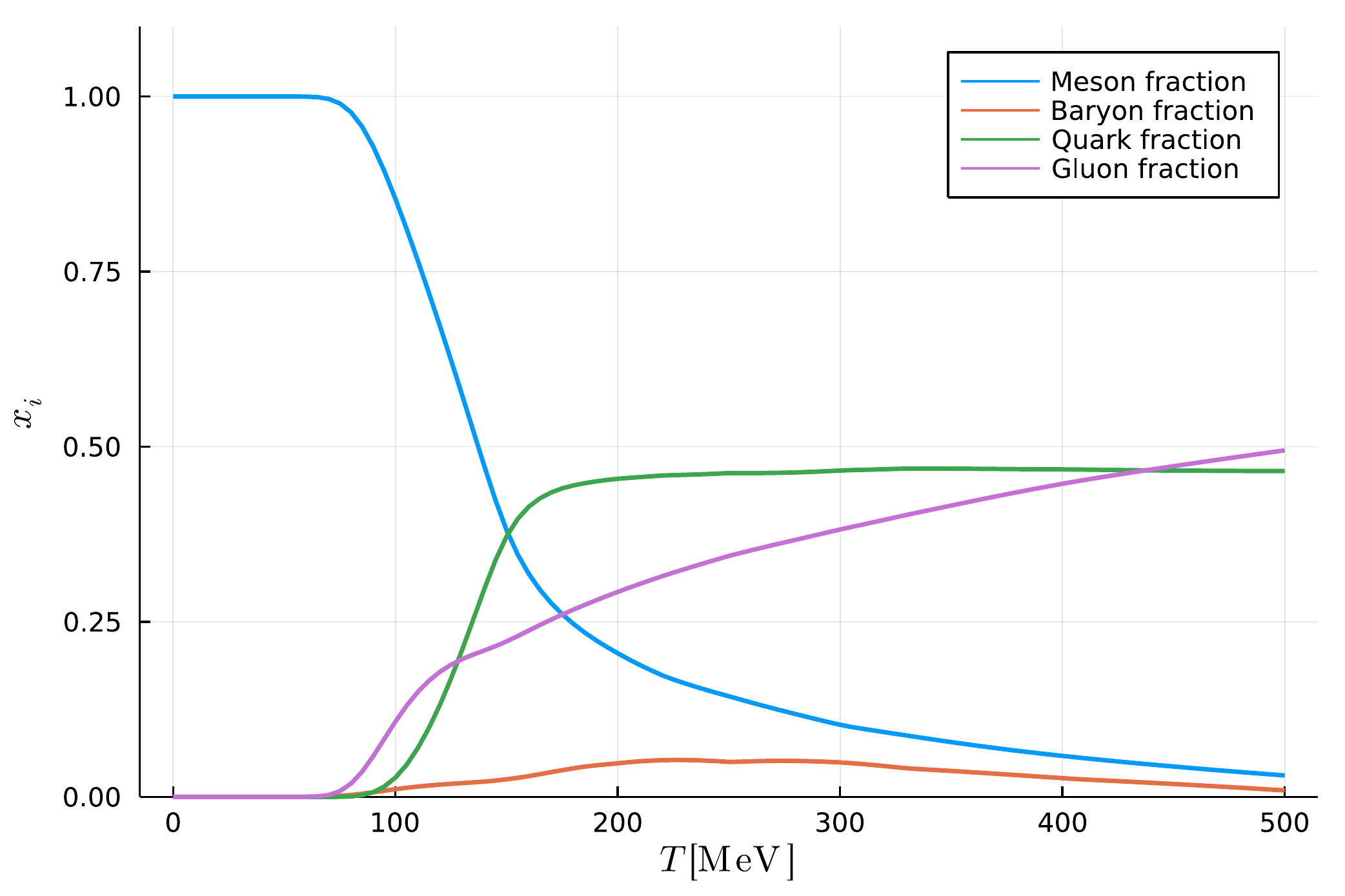}
	\caption{Fractions of particles in the system, evaluating their respective color scalar density.}
	\label{fig:fractions}
\end{figure}

In \cref{fig:thermodynamics} the thermodynamic quantities of the model, compared to the data of \gls{lqcd}, can be seen.
The fit was mainly done with the entropy density, while the other properties and the particle fraction has also been kept in mind.

\Cref{fig:masses} shows the effective mass $M_\pi$ of the pion, as lightest hadron, and the sum of two light quarks, representing the threshold mass of the pion $M^\mathrm{thr}_\pi$.
Further it shows the effective mass of the gluon $M_g$ for comparison.
The masses of partons diverge at low temperatures due to the confinement contribution ($n_\mathrm s^{-1/3}$).
At moderate temperatures of $T \approx [80 - 250]~\mathrm{MeV}$ the parton masses decrease and they can populate the system accordingly.
At higher temperatures of $T \gtrapprox 250~\mathrm{MeV}$ the linear contributions become dominant, which will become the leading term at asymptotic temperatures.
The structure at $T\approx 140~\mathrm{MeV}$, which is a local maximum for gluons and pions and a shoulder for light quarks, arises from the mixed terms in the rearrangement contributions, since there are no clear dominant terms around that temperature.
Up to this region the effective mass of pions is constant.

The particle fraction can be seen in \cref{fig:fractions}.
It is calculated as the ratio of the color scalar density of that species to the over-all color scalar density.
This means for the gluon fraction $x_g = n_{\mathrm s,g} / n_\mathrm s$, the quark fraction $x_q = \sum_{j = \{u,d,s\}} n_{\mathrm s,j} / n_\mathrm s$, the meson fraction  $x_m = \sum_{i \in \mathrm{mesons}} n_{\mathrm s,i} / n_\mathrm s$, and the baryon fraction  $x_b = \sum_{i \in \mathrm{baryons}} n_{\mathrm s,i}  / n_\mathrm s$ and the over-all sum results by definition in $1 = x_g + x_q + x_m + x_b$.
Up to a temperature $T\approx 80~\mathrm{MeV}$ mesons dominate the system.
Contributions of baryons stay over the entire temperature scale very small, due to their significantly higher masses, compared to mesons and the lack of chemical potential, which makes them dominant at high densities.
The first partons to populate the system are gluons, due to their initially smaller masses.
The correlation between effective mass and particle fraction can be clearly seen, as well as the non-trivial mixed term, which arises, when all species have similar fractions.

On a qualitative level, the model reproduces the thermodynamics very well.
Quantitatively, there are noticeable deviations, particularly in the transition area at $T\approx 150~\mathrm{MeV}$.
Also the inclination point, which would later define the phase diagram, is slighly different from the data, but here one needs to take into account, that the results from \gls{lqcd} are also subject to uncertainties.
The microscopic quantities of effective masses and particle fractions are model predictions, which can only be compared to predictions of other phenomenological models.
It is questionable, whether an onset of deconfined partons below $T = 100~\mathrm{MeV}$ is physical, but this behaviour can be explained by the simplicity of the model assumptions at this stage.
In particular, replacing the hadron phase shifts by experimentally observed scattering phase shifts \cite{Lo:2021oev} will suppress partonic contributions at lower temperatures.

Furthermore, it will be interesting to see whether the existence of partons at surprisingly low temperatures and hadron-like states in high temperatures will affect the susceptibilities and higher-order fluctuations and force the use of more refined phase shifts. Inclusion of the vector self energies, and calculating these fluctuations is the next step in the development of this model.

As was discussed in \cite{Bastian:2018mmc}, cluster expansion model can be used to create an \gls{eos} with a critical endpoint when nucleons are the only hadronic degrees of freedom.
We have now generalised the model to include strangeness and the whole zoo of hadronic resonances, and shown that with even very crude choices of phase shifts it qualitatively reproduces the Lattice QCD results.
Thus cluster expansion model is a viable framework for constructing an Equation of State which covers the entire QCD phase diagram from low density--high temperature limit to high density-low temperature region, with great freedom to choose the location of the critical endpoint. Constructing such a "coast-to-coast" EoS is in our plans for near future.

\section*{Acknowledgement}
This work was supported by Polish National Science Center (NCN) under the grant No. 2019/32/C/ST2/00556 (N.-U.F.B.) and by the program Excellence Initiative--Research University of the University of Wroc\l{}aw of the Ministry of Education and Science (P.H.).

\appendix

\section{High temperature QCD in a quasiparticle picture}
\label{sec:laine}

In \cite{Laine:2006cp} higher order corrections to the asymptotic limit of \gls{qcd} are discussed.
Here we want to show that the leading terms can be expressed in a quasi-particle picture and therefore used to derive the functional form of self-energy shifts in our work.
The physical pressure of hot \gls{qcd} is written in the form
\begin{align}
	\frac{p_\mathrm{QCD}}{T^4} &\approx \alpha_\mathrm{E1}^\mathrm{\overline{MS}} + \hat g_3^2 \alpha_\mathrm{E2}^\mathrm{\overline{MS}}\,,
\end{align}
with $\hat g_3^2 = {g'}^2$, as we are only interested in the first order. The prime at the coupling factor $g'$ was added to not confuse it with the previously used in this work degeneracy factor $g$.
The $\alpha^\mathrm{\overline{MS}}$ are evaluated for $N_f=3$ flavours resulting in:
\begin{align}
	\alpha_\mathrm{E1}^\mathrm{\overline{MS}} &= d_A \frac{\pi^2}{45} + 4C_A \sum_{i=1}^{N_f} F_1 \left( \frac{m_i^2}{T^2}, \frac{\mu_i}{T}\right)\\
\begin{split}
	\alpha_\mathrm{E2}^\mathrm{\overline{MS}} &= - \frac{d_A C_A}{144} - d_A \sum_{i=1}^{N_f} \left\{\frac 16 F_2 \left( \frac{m_i^2}{T^2}, \frac{\mu_i}{T}\right) \left[1 + 6 F_2 \left( \frac{m_i^2}{T^2}, \frac{\mu_i}{T}\right)\right]\right.\\
											&\qquad\qquad\left.+ \frac{m_i^2}{4\pi^2T^2} \left(3\ln\frac{\bar\mu}{m_i} + 2\right) F_2 \left( \frac{m_i^2}{T^2}, \frac{\mu_i}{T}\right) - \frac{2m_i^2}{T^2} F_4 \left( \frac{m_i^2}{T^2}, \frac{\mu_i}{T}\right)\right\}
\end{split}
\end{align}
Extracting the leading orders at asymptotically high temperatures gives us a constant term, which represents the Stefan-Boltzmann result for gluons followed by the expression:
\begin{align}
	\frac{p_\mathrm{QCD}}{T^4} &\stackrel{T\rightarrow\infty}{\sim}
	    4C_A \sum_{i=1}^{N_f} F_1 \left( \frac{m_i^2}{T^2}, \frac{\mu_i}{T}\right)
	- {g'}^2 d_A \sum_{i=1}^{N_f} \frac 16 F_2 \left( \frac{m_i^2}{T^2}, \frac{\mu_i}{T}\right) \left[1 + 6 F_2 \left( \frac{m_i^2}{T^2}, \frac{\mu_i}{T}\right)\right]\,.\label{eqn:laine_leading}
\end{align}
The thermal functions $F_{1}$ and $F_{2}$ are defined as:
\begin{subequations}
\begin{align}
	F_1 (y, \hat\mu) &= \frac{1}{24\pi^2} \int_0^\infty \mathrm dx x \sqrt{\frac{x}{x + y}}
	    \left(\frac{1}{\mathrm e^{\sqrt{x+y} - \hat\mu} + 1} + \frac{1}{\mathrm e^{\sqrt{x+y} - \hat\mu} + 1}\right)\\
	F_2 (y, \hat\mu) &= \frac{1}{24\pi^2} \int_0^\infty \mathrm dx \sqrt{\frac{x}{x + y}}
	    \left(\frac{1}{\mathrm e^{\sqrt{x+y} - \hat\mu} + 1} + \frac{1}{\mathrm e^{\sqrt{x+y} - \hat\mu} + 1}\right)\, .
\end{align}
\end{subequations}
Using the substitution $x = p^2/T^2$ one gets
\begin{align}
	F_1 (\frac{m^2}{T^2}, \frac{\mu}{T}) &= \frac{1}{12\pi^2 T^4} \int_0^\infty \mathrm dp p^2 \frac{p^2}{\sqrt{p^2 + m^2}}
	    \left(f + \bar f\right)
	= \frac{1}{2gT^4} p^\mathrm{id} (T, g, m, \mu)\,,
\end{align}
which can be recognized as the ideal gas pressure $p^\mathrm{id}$ of particles with mass $m$, the degeneracy factor $g$, temperature $T$ and chemical potential $\mu$.
Considering, that the mass $m$ is not constant, but rather has a logarithmic temperature dependence, this term can be treated as a quasi particle expression.
Utilizing the same substitution, the expression for $F_2$ takes the form
\begin{align}
	F_2 (\frac{m^2}{T^2}, \frac{\mu}{T}) &= \frac{1}{12\pi^2 T^2} \int_0^\infty \mathrm dp p^2 \frac{1}{\sqrt{p^2 + m^2}}
	    \left(f + \bar f\right)
	= \frac{1}{2g m T^2} n_\mathrm s^\mathrm{id} (T, g, m, \mu)\,,
\end{align}
which results in the scalar density $n_\mathrm s^\mathrm{id}$ of the same ideal gas.

Now we can write the leading terms of \eqref{eqn:laine_leading} as
\begin{align}
    p_\mathrm{QCD} &\stackrel{T\rightarrow\infty}{=\joinrel=}
	    \sum_{i=1}^{N_f} p^\mathrm{id} (T, g_i, m_i, \mu_i)
	+ A \sum_{i=1}^{N_f} n_\mathrm s^\mathrm{id} (T, g_i, m_i, \mu_i) 
	+ B \sum_{i=1}^{N_f} \left(n_\mathrm s^\mathrm{id} (T, g_i, m_i, \mu_i) \right)^2.
\end{align}
All non-vanishing coefficients are absorbed in the coefficients $A$ and $B$, whose exact value is not relevant for the schematic discussion here.
It should be noted, that the temperature dependence of the coupling constant $g'$ and the mass $m$ compensate in $B$, but slightly remain in $A$, which causes non-leading effects.

\printbibliography


\end{document}

%% file: niels-glossary.tex
\newabbreviation[longplural={equations of state},description={Equations of state relate thermodynamic quantities, which describe the state of matter.}]{eos}{EoS}{equation of state}
\newabbreviation[shortplural={HIC}]{hic}{HIC}{heavy-ion collisions}
\newabbreviation[description={Nuclotron-based Ion Collider fAcility in Dubna, Russia}]{nica}{NICA}{Nuclotron-based Ion Collider fAcility}
\newabbreviation{qmd}{QMD}{quantum molecular dynamic}
\newabbreviation{urqmd}{UrQMD}{ultra-relativistic quantum molecular dynamic}
\newabbreviation{amd}{AMD}{antisymmetrized molecular dynamics}
\newabbreviation{rdf}{RDF}{relativistic density functional}
\newabbreviation{sfm}{SFM}{string-flip model}
\newabbreviation{qcd}{QCD}{quantum chromodynamics}
\newabbreviation{qgp}{QGP}{quark-gluon plasma}
\newabbreviation{cep}{CEP}{critical endpoint}
\newabbreviation{njl}{NJL}{Nambu-Jona-Lasinio}
	\newabbreviation[parent=njl,description={}]{hnjl}{hNJL}{higher order \glsentryshort{njl}}
	\newabbreviation[parent=njl,description={}]{hnjlb}{hNJLb}{higher order \glsentryshort{njl} with bag constant}
	\newabbreviation[parent=njl,description={}]{pnjl}{PNJL}{Polyakov-loop \glsentrylong{njl}}
\newabbreviation{nse}{NSE}{nuclear statistical equilibrium}
\newabbreviation{cve}{CVE}{cluster virial expansion}
\newabbreviation{bnsm}{BNSM}{binary neutron star merger}
\newabbreviation{tov}{TOV}{Tolman--Oppenheimer--Volkoff}
\newabbreviation{gr}{GR}{general relativity}
\newabbreviation[longplural={core-collapse supernovae}]{ccsn}{CCSN}{core-collapse supernova}
\newabbreviation[description={Zero Age Main Sequence is the time when a star starts burning hydrogen and therefore joins the main sequence on the Hertzsprung-Russell diagram}]{zams}{ZAMS}{zero-age main sequence}
\newabbreviation{sb}{SB}{Stefan-Boltzmann}

\newabbreviation{n3lo}{N$^3$LO}{next-to-next-to-next-to-leading order}
\newabbreviation{ceft}{$\chi$EFT}{chiral effective field theory}

\newabbreviation{hrg}{HRG}{hadron resonance gas}
\newabbreviation{htl}{HTL}{hard-thermal loop}
\newabbreviation{pdg}{PDG}{particle data group}

\newglossaryentry{bu}{name={Beth-Uhlenbeck},description={}}
	\newglossaryentry{gbu}{name={generalized Beth-Uhlenbeck},parent=bu,description={}}
\newglossaryentry{cern}{name={CERN},description={European Organization for Nuclear Research in Geneva, Switzerland}}
\newglossaryentry{lhc}{name={LHC},description={Large Hadron Collider experiment at \gls{cern}}}
\newglossaryentry{alice}{name={ALICE},description={A Large Ion Collider Experiment}}
\newglossaryentry{rhic}{name={RHIC},description={Relativistic Heavy Ion Collider at the Brookhaven National Laboratory in Upton, USA}}
\newglossaryentry{fair}{name={FAIR},description={Facility for Antiproton and Ion Research at GSI in Darmstadt, Germany}}
\newglossaryentry{dd2all}{name={DD2(F)},description={Walecka-type hadron model with density dependent couplings}}
	\newglossaryentry{dd2}{name={DD2},parent=dd2all,description={}}
	\newglossaryentry{dd2f}{name={DD2F},parent=dd2all,description={}}
\newglossaryentry{theseus}{name={THESEUS},description={Three-fluid Hydrodynamics-based Event Simulator Extended by UrQMD final State interactions}}
\newglossaryentry{rhic:bes}{name={RHIC BES},parent=rhic,description={}}
\newglossaryentry{rhic:fxt}{name={RHIC FXT},parent=rhic,description={}}
\newglossaryentry{na61}{name={NA61},description={NA61}}
\newglossaryentry{lqcd}{name={Lattice QCD},description={lattice QCD}}
\newglossaryentry{pqcd}{name={pQCD},description={perturbative QCD}}
\newglossaryentry{nicer}{name={NICER}, description={Neutron Star Interior Composition Explorer \url{https://www.nasa.gov/nicer}}}